# Low temperature high-frequency monolayer graphene conductivity as a manifestation of Zitterbewegung

Natalie E. Firsova and Sergey A. Ktitorov,

Ioffe Institute, St. Petersburg 195267, Russia

**Abstract**

We consider the Zitterbewegung of Dirac electrons in the monolayer graphene as the non-relativistic analog of the phenomenon predicted by E. Schrödinger for the relativistic electrons in the free space. So we show that the Dirac electrons of monolayer graphene oscillate, i.e. produce fast fluctuation of an electron position around the mean value with rather high frequency but much less than the relativistic electrons in the free space. We also study relation between the Zitterbewegung of the conductivity electrons wave packet formed by the Fermi-Dirac distribution and the low temperature high-frequency complex conductivity of the monolayer graphene. Thus we find that the electromagnetic resonance properties of the monolayer graphene can be simulated by the set of the equivalent oscillatory circuits. This is useful in order to impart illustrativeness to electromagnetic processes that is particularly important for incorporation of graphene into electronic systems.

PACS 73.22.Pr, 85.35.-p, 74.25.N-

## 1. Introduction

Unique features of electronic states in graphene attract a permanent attention of theorists. In addition to such distinguished characteristics as colossal electronic mobility and gigantic thermoconductivity it is interesting to investigate the quantum electrodynamics effects based on the mathematical similarity of the relativistic states in the high energy physics and solid state physics. One of such effects is the intriguing Zitterbewegung (ZB) (or trembling motion from German) of the Dirac electrons. Here we will discuss this phenomenon in the monolayer graphene with special emphasis on relation to the low temperature high frequency conductivity problem.

## 2. Zitterbewegung of relativistic electrons in free space.

ZB is a fast oscillating motion of elementary particles, in particular electrons that obey the Dirac equation. ZB phenomenon was first predicted by Erwin Schrödinger in 1930 [1] as a result of his analysis of the wave packet solutions of the Dirac equation for electrons in free space. He concluded that the interference between positive and negative energy states produced what appeared to be a fluctuation (at the speed of light) of the position of an electron around the mean value with the frequency $2mc^2/\hbar$, where $m$ is the electron mass, $c$ is the speed of light. One can obtain this prediction in the following way.

The velocity operator is determined by the commutator:

$$\frac{d\hat{x}_j}{dt} = \frac{i}{\hbar}[\hat{H},\ \hat{x}_j] = -\frac{c}{\hbar}[\gamma_0\gamma_j\hat{p}_j,\ \hat{x}_j] = ic\gamma_0\gamma_j = \hat{v}_j \qquad (1)$$

The Heisenberg equation for the velocity operator evolution reads

$$\frac{d\hat{v}_j}{dt} = \frac{i}{\hbar}[\hat{H},\ \hat{v}_j] = \frac{i}{\hbar}(-2\hat{H}\hat{v}_j + \{\hat{H},\ \hat{v}_j\}) = \frac{i}{\hbar}(-2\hat{H}\hat{v}_j + 2c^2 p_j) \qquad (2)$$

A solution of this equation reads:

$$\hat{v}_j(t) = c^2\,\hat{p}_j/E + \big(\hat{v}_j(0) - c^2\,\hat{p}_j/E\big)e^{-i2Et/\hbar}, \qquad (3)$$

where $E = \pm\sqrt{m^2c^4 + p^2c^2}$ is the electron energy. We have the following estimate for the minimal ZB frequency: $2mc^2/\hbar \approx 1.8 \times 10^{21}\ sec^{-1}$. ZB of a free relativistic particle has never been observed because of the huge value of the trembling frequency.

### 3. Zitterbewegung in the monolayer graphene.

We study ZB in the monolayer graphene. For this purpose we consider the Heisenberg evolution equation

$$d\hat{V}/dt = (i/\hbar)[\hat{H}_0, \hat{V}]. \tag{4}$$

where $\hat{H}_0$ is the free Dirac Hamiltonian in the monolayer graphene,

$$\hat{H}_0(\mathbf{k}) = \hbar v_F \mathbf{k} \cdot \boldsymbol{\sigma} = \hbar v_F (k_1\hat{\sigma}_1 + k_2\hat{\sigma}_2) = \hbar v_F k \begin{pmatrix} 0 & e^{-i\varphi} \\ e^{i\varphi} & 0 \end{pmatrix}, \tag{5}$$

$$\hat{\sigma}_1 = \begin{pmatrix} 0 & 1 \\ 1 & 0 \end{pmatrix}, \quad \hat{\sigma}_2 = \begin{pmatrix} 0 & -i \\ i & 0 \end{pmatrix}, \quad \mathbf{k} = (k_1, k_2), \quad k_1 = k \cos\varphi, \quad k_2 = k \sin\varphi. \tag{6}$$

Then

$$\hat{V}(t) = e^{(i/\hbar)\hat{H}_0 t}\hat{V}(0)e^{-(i/\hbar)\hat{H}_0 t}, \tag{7}$$

where

$$\hat{V}(0) = v_F\hat{\sigma}_1 = \begin{pmatrix} 0 & v_F \\ v_F & 0 \end{pmatrix}. \tag{8}$$

as far as

$$\hat{H}_0 = \hbar v_F k \begin{pmatrix} 0 & e^{-i\varphi} \\ e^{i\varphi} & 0 \end{pmatrix} = \hbar v_F k \hat{U}^{-1} \begin{pmatrix} 1 & 0 \\ 0 & -1 \end{pmatrix} \hat{U}, \tag{9}$$

where

$$\hat{U} = \frac{1}{\sqrt{2}}\begin{pmatrix} e^{i\varphi} & 1 \\ 1 & -e^{-i\varphi} \end{pmatrix}, \quad \hat{U}^{-1} = \frac{1}{\sqrt{2}}\begin{pmatrix} e^{-i\varphi} & 1 \\ 1 & -e^{i\varphi} \end{pmatrix}, \tag{10}$$

we have

$$e^{\pm(i/\hbar)\hat{H}_0 t} = \hat{U}^{-1}\begin{pmatrix} e^{\pm iv_F k t} & 0 \\ 0 & e^{\mp iv_F k t} \end{pmatrix}\hat{U}. \tag{11}$$

Hence we obtain

$$\hat{V}(t) = v_F\hat{U}^{-1}\begin{pmatrix} \cos\varphi & -i\sin\varphi\, e^{i\varphi+2iv_F kt} \\ i\sin\varphi\, e^{-i\varphi-2iv_F kt} & -\cos\varphi \end{pmatrix}\hat{U}. \tag{12}$$

Thus we see that the electron with a given energy $E = \hbar k v_F$ oscillates with the frequency

$$\omega_{ZB}(k) = 2v_F k. \tag{13}$$

Similarly to the case of the relativistic electrons in the free space, the ZB phenomenon in graphene is related to the uncertainty of the particle coordinates due to unavoidable creation of the electron-hole pairs during the process of measurement [2]. On the other hand, the ZB can be considered as a sort of inter-band transitions accompanied by creation of virtual electron-hole pairs. Notice that the unitary operator $U$ used on [5] differs from one in [2].

## 4. Kramers – Krönig relations and low – frequency behavior of conductivity.

Let us write a formula for the high-frequency conductivity [3, 4, 5]

$$\sigma(\omega) = \frac{e^2}{4\hbar}\theta(\hbar\omega - 2E_F) + \frac{e^2}{\hbar}\frac{E_F}{\hbar}\delta(\omega) + i\frac{1}{\pi}\frac{e^2}{\hbar}\left[\frac{E_F}{\hbar\omega} + \frac{1}{4}\log\left|\frac{2E_F - \hbar\omega}{2E_F + \hbar\omega}\right|\right], \quad (14)$$

where $\theta(x)$ is the Heaviside step function. This formula is valid for high enough frequencies $\omega\tau \gg 1$, where $\tau^{-1}$ is the scattering frequency. Since for low enough temperature we have an estimate $\tau \sim 10^{-12} sec$, the formula works for frequencies higher, than a few THz.

One can see that the real and imaginary parts of the conductivity satisfy the Kramers – Krönig dispersion relations [6]:

$$Re\sigma(\omega) = \frac{2}{\pi} P \int_0^\infty ds \frac{s\, Im\sigma(s)}{s^2 - \omega^2}, \quad (15)$$

$$Im\sigma(\omega) = -\frac{2\omega}{\pi} P \int_0^\infty ds \frac{s\, Re\sigma(s)}{s^2 - \omega^2} \quad (16)$$

Let us investigate a possibility to simulate the low-temperature reactive conductivity of the monolayer graphene with a use of an equivalent circuit. It was shown in [5] that for low frequencies lying in the terahertz band, i.e. near the Dirac point, when

$$\omega/(2k_F v_F) \ll 1, \quad (17)$$

the conductivity formula can be rewritten in the form

$$\sigma(\omega) \approx (i\omega L(k_F))^{-1} + i\omega C(k_F). \quad (18)$$

where

$$L(k) = \frac{\hbar}{e^2}\frac{\pi}{v_F k} = \frac{2\hbar}{e^2}\frac{\pi}{\omega_{ZB}(k)}, \quad C(k) = \frac{1}{4\pi}\frac{e^2}{\hbar}\frac{1}{v_F k} = \frac{1}{2\pi}\frac{e^2}{\hbar}\frac{1}{\omega_{ZB}(k)}. \quad (19)$$

and $k_F$ and $v_F$ are respectively the Fermi wave number and velocity. Thus the following formulae for graphene inductivity and capacity were obtained in [5]

$$L(k_F) = (2k_F v_F \sigma_0)^{-1} = \hbar/(\pi e^2 k_F v_F), \quad (20)$$

$$C(k_F) = \frac{\sigma_0}{2k_F v_F} = \frac{\pi e^2}{4\hbar k_F v_F} = \frac{\pi e^2}{4 E_F}. \quad (21)$$

It is necessary to note that similar expressions for the inductance and capacitance were obtained on the physical grounds in [7, 8] (apart from the numerical factor in the case of the capacitance). So there were found in [5] the formulae (20), (21) for monolayer graphene inductivity and capacity. Notice that in opposite to [7, 8] the derivation made in [5] is based on the unified approach.

To obtain this result (18) one should write the imaginary part of (14) in the form:

$$Im\sigma(\omega) = i\sigma_0\left[\frac{k_F v_F}{\omega} + \frac{1}{4}\log\left|\frac{1 - \frac{\omega}{2k_F v_F}}{1 + \frac{\omega}{2k_F v_F}}\right|\right], \qquad \sigma_0 = \frac{1}{\pi}\frac{e^2}{\hbar}. \quad (22)$$

Thus, the graphene conductivity was simulated in [5] in the low – frequency limit near the Dirac point by the equivalent parallel oscillatory circuit with the inductance $L(k_F)$ and capacitance $C(k_F)$.

This simulation and the formulae (20), (21) for monolayer graphene inductivity and capacity are valid under condition (17) i. e. near Dirac point that means that it works for frequencies larger, than $\tau^{-1} \sim 10^{-12} sec^{-1}$ i. e. a few THz and when $\omega << \omega_{ZB}(k_F) = 2k_F v_F = 2\sqrt{2\pi(N/S)} v_F$. Assuming typical magnitude $n = N/S = 10^{11} cm^{-2}$ for the 2-d electronic gas density we obtain the estimate $\omega_{ZB}(k_F) \sim 15$ THz.

Note that in [7] the suggestion was made that electrons in graphene must exhibit a nonzero mass, when collectively excited. Using this notion the inertial accelerating of the electron collective mass and phase delay of the resulting current were considered. On the basis of this model the so-called kinetic inductance representing the reluctance of the collective mass to acceleration was introduced, calculated and measured. The found in [7] expression for inductance coincides with the obtained in [5] formula (20) for unity width. Analyzing this formula for inductivity we see that $L \sim n^{-1/2}$. Just such dependence was observed in the experiment in [7].

The formula for quantum capacitance based on the two-dimensional free electron gas model was suggested in [8]. The expression for capacitance obtained there coincides with the found in [5] expression (21) being applied to the unity square, i.e. with

$$c = Cn = \sqrt{\pi n}\, e^2/(4\hbar v_F),$$

up to the numerical factor $8/\pi$.

## 5. Equivalent circuit for the graphene conductivity and Zitterbewegung in the monolayer graphene.

Let us consider now the simulation problem for the case of large frequency $\omega \geq 2v_F k_F = \omega_{ZB}(k_F)$. Notice that

$$\log \left| \frac{k_F - \frac{\omega}{2v_F}}{k_F + \frac{\omega}{2v_F}} \right| = \frac{\omega}{v_F} \int_{k_F}^{\infty} \frac{dk}{\left(\frac{\omega}{2v_F}\right)^2 - k^2}. \qquad (23)$$

Note that expansion of the formula (23) in powers of $x$ gives the following series [9]:

$$\log \frac{1 - \frac{\omega}{2k_F v_F}}{1 + \frac{\omega}{2k_F v_F}} = -2 \sum_{n=1}^{\infty} \frac{1}{2n-1}\left(\frac{\omega}{2k_F v_F}\right)^{2n-1} \quad \text{for } \left(\frac{\omega}{2k_F v_F}\right)^2 < 1, \qquad (24)$$

$$\log \frac{1 - \frac{\omega}{2k_F v_F}}{1 + \frac{\omega}{2k_F v_F}} = -2 \sum_{n=1}^{\infty} \frac{1}{2n-1}\left(\frac{\omega}{2k_F v_F}\right)^{-2n+1} \quad \text{for } \left(\frac{\omega}{2k_F v_F}\right)^2 > 1. \qquad (25)$$

Using (23) we have

$$Im\sigma(\omega) = -\frac{1}{i\omega L(k_F)} - \frac{i}{2\pi} \frac{e^2}{\hbar} \frac{\omega}{2v_F} \int_{k_F}^{\infty} \frac{dk}{k^2 - \left(\frac{\omega}{2v_F}\right)^2} = -\frac{1}{i\omega L(k_F)} - \int_{k_F}^{\infty} \sigma_*(k, \omega) \frac{dk}{k}, \qquad (26)$$

where

$$\sigma_*(k, \omega) = \frac{i\left(\frac{\omega}{\omega_{ZB}(k)}\right)}{1 - \left(\frac{\omega}{\omega_{ZB}(k)}\right)^2} \sqrt{\frac{C(k)}{L(k)}}, \qquad (27)$$

and $\omega_{ZB}(k) = 2v_F k$ (see (13)). As soon as (see (19))

$$(C(k)/L(k))^{1/2} = e^2/(2\pi\hbar), \qquad (28)$$

we have

$$\sigma_*(k,\omega) = \frac{1}{2\pi}\frac{e^2}{\hbar}\frac{i\left(\frac{\omega}{\omega_{ZB}(k)}\right)}{1-\left(\frac{\omega}{\omega_{ZB}(k)}\right)^2}. \tag{29}$$

Let us analyze now the formula for conductivity (22). For this aim we consider a series oscillatory circuit with the inductivity $L(k)$ and capacity $C(k)$. We define its conductivity by $\tilde{\sigma}(k,\omega)$. Then

$$\tilde{\sigma}(k,\omega) = (i\omega L(k) + 1/(i\omega C(k)))^{-1}. \tag{30}$$

Hence

$$\tilde{\sigma}(k,\omega) = \frac{i\left(\frac{\omega}{\omega_{ZB}(k)}\right)}{1-\left(\frac{\omega}{\omega_{ZB}(k)}\right)^2}\sqrt{\frac{C(k)}{L(k)}} = \frac{1}{2\pi}\frac{e^2}{\hbar}\frac{i\left(\frac{\omega}{\omega_{ZB}(k)}\right)}{1-\left(\frac{\omega}{\omega_{ZB}(k)}\right)^2}, \tag{31}$$

Notice that the resonance frequency of this series oscillatory circuit at the given $k$ equals to the ZB frequency $\omega_{ZB}(k) = 2v_F k$. Thus we can write (see (29)):

$$\tilde{\sigma}(k,\omega) = \sigma_*(k,\omega) = \frac{i\left(\frac{\omega}{\omega_{ZB}(k)}\right)}{1-\left(\frac{\omega}{\omega_{ZB}(k)}\right)^2}\sqrt{\frac{C(k)}{L(k)}} = \frac{1}{2\pi}\frac{e^2}{\hbar}\frac{i\left(\frac{\omega}{\omega_{ZB}(k)}\right)}{1-\left(\frac{\omega}{\omega_{ZB}(k)}\right)^2}. \tag{32}$$

So we have

$$Im\sigma(\omega) = -\frac{1}{i\omega L(k_F)} - \frac{i}{2\pi}\frac{e^2}{\hbar}\frac{\omega}{2v_F}\int_{k_F}^{\infty}\frac{dk}{k^2-\left(\frac{\omega}{2v_F}\right)^2} = -\frac{1}{i\omega L(k_F)} - \int_{k_F}^{\infty}\tilde{\sigma}(k,\omega)\frac{dk}{k} \tag{33}$$

Formula (33) assumes a transparent physical picture: the first term is the Drude inductive reactance, the second term presents the parallel connected to this inductance set of parallel connected series oscillators, contribution of which is determined by the weight function $1/k$. Thus, we have an infinite network comprising an inductivity $L(k_F)$ and an infinite continuum number of oscillators (the series oscillatory circuits) with inductivity $L(k)$ and capacity $C(k)$.

If $\omega \geq 2v_F k_F$, i. e. above the threshold, then there is definitely an oscillator with the resonance frequency $\omega_{ZB}(k)$ coinciding with $\omega$. As a result of this resonance, a reactive resistance disappears. The active resistivity value is given by formula (14): $\frac{4\hbar}{e^2}$. So the nature of this active resistance appearing above the threshold is ZB. Note that the integrand in (33) contains the density of oscillators $1/k$:

$$2\frac{dk}{k} = \frac{2\pi k dk}{\pi k^2}, \tag{34}$$

i.e. the ratio of the area of the ring between circles with radii $k$ and $k + dk$ to the area of the circle of radius $k$. For more illustrativeness we can substitute the integral by the proper integral sum with summing over the discrete values of the wave vector $k_n = 2\pi n/B$ with $B$ being the linear size of the sample. This infinite sum can be truncated introducing the cut-off wave vector. Thus we come to the finite-element equivalent circuit presented in the Fig. 1.

Note that the lower possible value of the predicted by Schrödinger ZB frequency $2mc^2/\hbar \approx 1.8 \times 10^{21}\ sec^{-1}$ is $10^8$ times larger, than the estimated above minimal ZB in graphene. This promises using graphene as a model of some quantum electrodynamics processes.

## 6. Conclusions

We have studied Zitterbewegung of the Dirac electrons in the monolayer graphene, which proved to be the non-relativistic analog of the corresponding phenomenon predicted by E. Schrödinger for the relativistic electrons in the free space. So we have shown that the Dirac electrons of the monolayer graphene oscillate, i. e. produce fast fluctuations of an electron position around the mean value with a high frequency. The ZB frequency in graphene was found to be $10^8$ times lower, than in the space.

A relation between the ZB of the conductivity electrons wave packet formed by the Fermi-Dirac distribution and the high-frequency complex conductivity of the monolayer graphene is established. We have shown that the analyzed formula for the low temperature high-frequency monolayer graphene complex conductivity assumes a transparent physical picture: the first term is the Drude inductive reactance, the second term presents the parallel connected to this inductance set of (parallel connected) series $LC$-oscillators with $k \geq k_F$, relative contribution of which is determined by the weight function $1/k$. Thus the high – frequency behavior is described by an infinite network comprising the inductivity $L(k_F)$ and an infinite continuum number of oscillators (the series oscillatory circuits) with inductivity $L(k)$ and capacity $C(k)$, parameterized by the continuum of wave vectors. For frequencies, lying within the range $\tau^{-1} \ll \omega \ll 2k_F v_F$, i. e. for frequencies larger, than a few THz and less than about 15 THz, as it was shown in [5] the equivalent circuit is reduced to a single parallel oscillatory circuit with inductivity $L(k_F)$ and capacity $C(k_F)$ (see (20, 21). Notice, that this frequency range is a very interesting one for practice.

If $\omega \geq 2v_F k_F$, i. e. above the threshold ($2v_F k_F$ is equal about 15 THz), there exists definitely at least one oscillator with the resonance frequency $\omega_{ZB}(k) = 2v_F k$ coinciding with $\omega$. As a result of this resonance, an active resistance $R = 16\ k\Omega$ appears in the monolayer graphene for $k \geq k_F$. So the nature of the active resistance of the monolayer graphene is ZB. For finite samples the equivalent circuit is presented in Fig.1.

Thus we have found that the electromagnetic resonance properties of the monolayer graphene can be simulated by the set of the equivalent oscillatory circuits that can help to construct electronic systems including graphene elements.

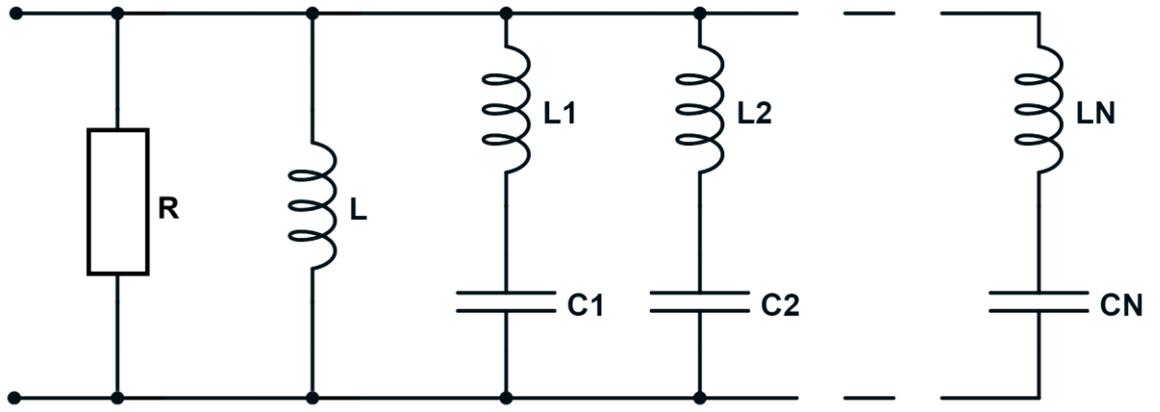

Fig, 1. Equivalent circuit